\documentclass[prd,superscriptaddress,amsfonts,amssymb,onecolumn,amsmath,showpacs]{revtex4-2}
\usepackage{bm}
\usepackage{amsfonts}
\usepackage{tensor}
\newcommand{\lc}[1]{\overset{\circ}{#1}}

\usepackage{latexsym}
\usepackage[utf8]{inputenc}
\usepackage{graphicx}
\usepackage{amsmath}
\usepackage{palatino}
\usepackage{mathpazo}
\usepackage[british]{babel}
\usepackage{hhline}
\usepackage{multirow}
\usepackage{textcomp}
\usepackage{float}
\usepackage{booktabs}
\usepackage{dcolumn}
\usepackage{lipsum} 
\usepackage{mdframed}
\usepackage[]{mdframed}
\usepackage{hhline}
\usepackage{multirow}
\usepackage{ragged2e}
\usepackage{hyperref}
\hypersetup{colorlinks,citecolor=blue}
\hypersetup{colorlinks=true,linkcolor=red,filecolor=magenta,    urlcolor=cyan}
\usepackage{amsmath}
\usepackage{xcolor}
\usepackage{orcidlink}
\usepackage{epsfig}
\usepackage{caption}
\usepackage{subcaption}
\usepackage{commath}

\def\jnl@style{\it}
\def\aaref@jnl#1{{\jnl@style#1}}

\def\aaref@jnl#1{{\jnl@style#1}}

\def\aj{\aaref@jnl{AJ}}                   
\def\apj{\aaref@jnl{ApJ}}                 
\def\apjl{\aaref@jnl{ApJ}}                
\def\apjs{\aaref@jnl{ApJS}}               
\def\apss{\aaref@jnl{Ap\&SS}}             
\def\aap{\aaref@jnl{A\&A}}                
\def\aapr{\aaref@jnl{A\&A~Rev.}}          
\def\aaps{\aaref@jnl{A\&AS}}              
\def\mnras{\aaref@jnl{Mon.~Not.~Roy.~Astron.~Soc.}}             
\def\prd{\aaref@jnl{Phys.~Rev.~D}}        
\def\prc{\aaref@jnl{Phys.~Rev.~C}}  
\def\prl{\aaref@jnl{Phys.~Rev.~Lett.}}    
\def\qjras{\aaref@jnl{QJRAS}}             
\def\skytel{\aaref@jnl{S\&T}}             
\def\ssr{\aaref@jnl{Space~Sci.~Rev.}}     
\def\zap{\aaref@jnl{ZAp}}                 
\def\nat{\aaref@jnl{Nature}}              
\def\aplett{\aaref@jnl{Astrophys.~Lett.}} 
\def\apspr{\aaref@jnl{Astrophys.~Space~Phys.~Res.}} 
\def\physrep{\aaref@jnl{Phys.~Rep.}}      
\def\physscr{\aaref@jnl{Phys.~Scr}}       
\def\commat{\aaref@jnl{Comm.~Math.~Phys.}}              
\def\science{\aaref@jnl{Science}}               
\def\cqg{\aaref@jnl{Classical Quant.~Grav.}}            
\def\jpcs{\aaref@jnl{JPCS}}                                     
\def\ijmpd{\aaref@jnl{Int.~J.~Mod.~Phys.~D}}                    
\def\grg{\aaref@jnl{Gen.~Relat.~Gravit.}}               
\def\rpp{\aaref@jnl{Rep.~Prog.~Phys.}}          
\def\npa{\aaref@jnl{Nucl.~Phys.~A}}        
\def\lrr{\aaref@jnl{Living Rev.~Rel.}}                   
\def\jcap{\aaref@jnl{J.~Cosmology Astropart.~Phys.}}    
\def\rmp{\aaref@jnl{Rev.~Mod.~Phys.}}   
\def\epjc{\aaref@jnl{Eur.~Phys.~J.~C}} 
\def\plb{\aaref@jnl{~Phy.~Lett.~B}} 
\def\mpla{\aaref@jnl{Mod.~Phy.~Lett.~A}} 
\def\arxiv{\aaref@jnl{arxiv.org}}

\allowdisplaybreaks[1]

\begin{document}

\title{Dynamical Evolution in Generalized Scalar-Torsion Gravity with Extended Couplings}

\author{S. A. Kadam\orcidlink{0000-0002-2799-7870}}
\email{k.siddheshwar@sru.edu.in ;
\\k.siddheshwar47@gmail.com}
\affiliation{Department of Mathematics, School of CS $\&$ AI, SR University, Warangal, Telangana 506371, India}

\begin{abstract}
\textbf{Abstract:} In this study we explore the cosmological behavior of a non-minimally coupled scalar field that is linked to torsion gravity. We demonstrate the Sorkin-Schutz formalism with general power law teleparallel torsion coupling. The autonomous dynamical system has been formulated. The phase space diagrams have been analysed at each critical point. The critical points representing different eras of Universe evolution starting from radiation, dark matter (DM), and dark energy (DE) have been investigated. The scaling attractors with the viable range of model parameters have been obtained using exponential scalar field couplings. This modified version of the formalism describes some novel scaling solutions.  
\end{abstract}

\maketitle
\textbf{Keywords:} Sorkin-Schutz formalism, $f(T,\phi)$ gravity,   Autonomous Dynamical System Analysis
\section{Introduction}\label{Introduction}

The late-time acceleration of the Universe is one of the significant challenges in modern physics. This was first observed by surveys of type Ia supernovae \cite{Riess:1998cb,Perlmutter:1998np}, and this crucial finding has been supported by increasingly precise cosmological measurements, such as the cosmic microwave background (CMB) \cite{Hinshaw:2013,Planck:2015bue}, the Hubble constant \cite{PhysRevResearch.2.013028}, baryon acoustic oscillations (BAO) \cite{Favale:2024sdq}, and additional observations of type Ia supernovae \cite{BOSS:2012bus}. Despite the availability of observational data sets, there is a significant lack of theoretical understanding regarding the late-time acceleration. The standard cosmological model postulates the existence of a cosmological constant to account for this acceleration. However, this cosmological constant faces theoretical challenges due to its remarkably small observed value in comparison to predictions made by quantum field theory \cite{Lombriser:2019jia}. Moreover, this model has observational challenges, such as recent results from the Planck collaboration indicating a growing discrepancy between local and global estimates of $H0$ and $f_{\sigma 8}$ \cite{Aghanim:2018eyx}. It might also be essential to explore an alternative way to meet the growing requirements for establishing a workable theory of gravity. One possible way is to investigate teleparallel gravity (TG) theories \cite{bahamonde:2021teleparallel,Cai:2015emx,Clifton:2011jh}.\\ 

TG represents an alternative formulation of gravity known as teleparallel equivalent to general relativity (TEGR). In General Relativity (GR), we have a geometric theory founded on the Levi-Civita connection. This theory exhibits curvature but without
torsion. The parallel view to this is the formulation of TG, which operates using a connection termed the Weitzenböck connection \cite{Weitzenbock1923,Krssak:2015oua,Ferraro:2008ey}. Despite having a vanishing Ricci tensor, this connection features nontrivial torsion. This finding is utilized to create an action centered around a gravitational scalar known as the torsion scalar $T$, which is formulated using the contribution of the Weitzenböck connection. The dynamics of the action for TG are fully equivalent to those of GR. There are various methods to modify both GR and TG. These ways can show importance to get a better understanding of the universe's late-time acceleration without relying on a cosmological constant. One strategy involves modifying the gravitational sector, allowing for the consideration of both $f(R)$ and $f(T)$ theories of gravity \cite{Basilakos:2018arq,Ferraro:2006jd, Cai:2015emx,Bengochea:2008gz}. Generally, these lead to different dynamics, as it is no longer possible to express $f(R)$ as $f(T)$ plus a total derivative. $f(R)$ gravity typically results in fourth-order field equations \cite{Oikonomou:2020qah}. In a way, $f(T)$ theories are considered less problematic since this modification yields only second-order field equations with the drawback of loss of local Lorentz invariance. A recent paper \cite{Bahamonde:2015zma} explored a broader theory within the teleparallel framework known as $f(T,B)$ gravity, which $B$ represents a divergence of torsion, encompassing both $f(R)$ and $f(T)$ gravity as specific cases.\\

In order to make the analysis smoother while adding more degrees of freedom, one can examine scalar field models to describe the DM and DE \cite{Hussain_2024}. The scalar field models in which the scalar fields are coupled with torsion scalar $T$, such as those featuring a $\zeta \phi^{2} T$ term \cite{Bahamonde:2017ize}, also facilitate the exploration of modifications in TG \cite{Bajardi:2021tul} and were first discussed to study the nature of DE in \cite{Geng:2011aj}. These models provide additional degrees of freedom that could help stabilize high-energy fluctuations \cite{Cai:2015emx,Bahamonde:2017ize}.  Further modification of this formalism includes the addition of the terms of the form$ F(\phi)G(T)$ to the action, with $G(T)$ being a function of the torsion scalar $T$, or exploring a general function $f(T, \phi)$ \cite{Rodriguez_Benites_2025,Gonzalez-Espinoza:2020jss}. This modified gravity formalism is capable of studying different important phenomena of Universe evolution, such as inflation, cosmological singularities \cite{trivedi2023cosmological}, dynamical system analysis \cite{samaddar2023qualitative}, and the Noether symmetry approach \cite{Duchaniya_2023nor}. In this study our aim is to construct the autonomous dynamical system and to study its critical points, describing different epochs of the Universe evolution. We are demonstrating the Sorkin-Schutz formalism from Refs. \cite{128_SCHUT,129_Brown_1993,133_Amendola_2020} taking a model that is of the general power law form; the linear form of torsion scalar has been studied in Ref. \cite{Rodriguez_Benites_2025} and obtaining the constraints on the model parameter $n$. Moreover, this study aims to analyzes the scaling solutions obtained during the dynamcial analysis. The paper is organised as in Sec. \ref{TG}, we have presented the background equations that are required to formulate TG. In Sec. \ref{scalar_tensor_formalism} the scalar-tensor $f(T, \phi)$ gravity formalism has been presented with the detailed dynamical system analysis following subsections as, in Subsec. \ref{FLRWUniverse}, the required cosmological expressions have been presented with the field equations, and in Subsec. \ref{dynamical_system_analysis}, autonomous dynamical system analysis is presented.  In Subsec. \ref{critical_point_analyais}, the critical points have been discussed in detail. In Subsec. \ref{Numerical_results} the numerical results have been discussed. Finally, the conclusion is presented in Sec. \ref{conclusion}.

\section{TELEPARALLEL GRAVITY BASIC FORMULATION}\label{TG}
The TEGR, offers an alternative framework for understanding gravity in terms of torsion rather than curvature. TG is classified as a gauge theory associated with the translation group. The most effective method for developing teleparallel theories of gravity involves the tetrad ${e}^{A}_{\ \ \mu}$ (along with its inverses $E_{A}^{\ \ \mu}$) \cite{Bahamonde:2021gfp}, which serves as the fundamental variable of the theory in place of the metric, through the established relations
\begin{align}\label{metric_tetrad_rel}
    g_{\mu\nu} &= \tensor{e}{^A_\mu} \tensor{e}{^B_\nu} \eta_{AB}\,, & 
    \eta_{AB} &= \tensor{E}{_A^\mu} \tensor{E}{_B^\nu} g_{\mu\nu}\,.
\end{align}
In this context, Latin indices denote coordinates within the tangent space, whereas Greek indices still indicate indices on the overall manifold \cite{Cai:2015emx}. Additionally, \( g_{\mu\nu} \) represents the metric of spacetime, while \( \eta_{AB} = \text{diag}(-1, 1, 1, 1) \) denotes the Minkowski metric in tangent space. Like the metric, the tetrads need to meet orthogonality conditions represented in the following manner,
\begin{align}
    \tensor{e}{^A_\mu} \tensor{E}{_B^\mu} &= \delta^A_B\,, &  
    \tensor{e}{^A_\mu} \tensor{E}{_A^\nu} &= \delta^\nu_\mu\,.
\end{align}

The teleparallel connection can be explicitly described as \cite{Weitzenbock1923,Krssak:2015oua},
\begin{equation}
    \tensor{\Gamma}{^\sigma_{\nu\mu}} := \tensor{E}{_A^\sigma} 
    \left(\partial_{\mu} \tensor{e}{^A_\nu} + \tensor{\omega}{^A_{B\mu}} \tensor{e}{^B_\nu} \right)\,,
\end{equation}
where the spin connection of TG is characterized as
\begin{equation}
    {\omega^A}_{B\mu} = {\Lambda^A}_D(x) \partial_\mu {\Lambda_B}^D(x)\,.
\end{equation}
The components of a point-dependent local Lorentz transformation are denoted as ${\Lambda^A}_D(x)$. The contributions of the tetrad are enhanced by the flat spin connection $\tensor{\omega}{^A_{B\mu}}$, which plays a key role in integrating local Lorentz invariance into teleparallel theories. This is due to the explicit presence of Lorentz indices, which indicate the use of Lorentz frames. Just as tetrads are present in General Relativity (GR), spin connections are also included; however, a significant difference is that they are not flat in the context of GR \cite{misner1973gravitation}. The combination of tetrads and spin connections signifies the gravitational and local degrees of freedom in the equations of motion of Teleparallel Gravity (TG). In a similar manner to how the Levi-Civita connection leads to the Riemann tensor, the teleparallel connection directly results in the torsion tensor \cite{Hayashi:1979qx},
\begin{equation}
    {T^A}_{\mu\nu} = \partial_\mu {e^A}_\nu - \partial_\nu {e^A}_\mu + {\omega^A}_{B\mu} {e^B}_\nu - {\omega^A}_{B\nu} {e^B}_\mu,
\end{equation}
This tensor remains invariant under both diffeomorphisms and local Lorentz transformations. By suitably combining contractions of torsion tensors, it is possible to formulate a torsion scalar such that \cite{Krssak:2018ywd,Cai:2015emx,Aldrovandi:2013wha,Bahamonde:2021gfp},
\begin{equation}
    T := \frac{1}{4} \tensor{T}{^\alpha_{\mu\nu}} \tensor{T}{_\alpha^{\mu\nu}} 
    + \frac{1}{2} \tensor{T}{^\alpha_{\mu\nu}} \tensor{T}{^\nu^\mu_{\alpha}} 
    - \tensor{T}{^\alpha_{\mu\alpha}} \tensor{T}{^\beta^\mu_{\beta}} \,.
\end{equation}
which arises from a condition that $T$ must be equivalent to the curvature scalar $\tensor{\lc{R}}{}$ (up to a boundary term). In a manner similar to how the curvature scalar solely relies on the Levi-Civita connection, the torsion tensor is entirely dependent on the teleparallel connection.

By substituting the Levi-Civita connection with the teleparallel connection, measures of curvature invariably become zero, such as $R \equiv 0$ (where we highlight that $R = R(\tensor{\Gamma}{^\sigma_{\mu\nu}}) \quad \text{and} \quad 
\tensor{\lc{R}}{}$. In this framework, we can express the following relationship for the curvature and torsion scalars \cite{Bahamonde:2015zma,Farrugia:2016qqe}.

\begin{equation}\label{LC_TG_conn}
    R = \tensor{\lc{R}}{} + T - B = 0\,.
\end{equation}

In this context,  
\begin{equation}
B = \frac{2}{e} \partial_{\rho} \left( e \tensor[^\mu]{T}{^\mu_\rho} \right)\,,
\end{equation}
represents a total divergence term, where \( e = \det\left(\tensor{e}{^a_\mu}\right) = \sqrt{-g} \) denotes the determinant of the tetrad. This ensures that the general relativity (GR) and teleparallel equivalent of general relativity (TEGR) actions produce the same field equations.

The modified gravity models can be formulated alternatively based on either curvature or torsion theories, potentially resulting in different formalism. In the realm of modified teleparallel gravity, numerous studies have investigated DE and inflation influenced by scalar fields which are nonminimally coupled \cite{Geng:2011ka,Otalora:2013tba,Gonzalez-Espinoza:2020jss,Gonzalez_Espinoza_2020A,Gonzalezreconstruction2021}. Furthermore, research has been conducted on models that include nonlinear torsion terms, such as $f(T)$ gravity \cite{Bengochea:2008gz,Linder:2010py}. These torsion-based theories, in contrast to those based on curvature, have prompted significant investigation into cosmology at both early and late stages of Universe evolution \cite{Cai:2015emx}.
\section{ Interacting Scalar-Torsion Gravity in $f(T,\phi)$ Formalism}\label{scalar_tensor_formalism}

The action formula can be derived from the scalar-torsion gravity action $f(T,\phi)$ by including interactions that consider the transfer of energy and momentum between the scalar field and cold DM. This can be accomplished using the Sorkin-Schutz formalism \cite{128_SCHUT,129_Brown_1993,133_Amendola_2020,Rodriguez_Benites_2025}, as detailed below,
\begin{equation}
    S = \int d^4x e [f(T, \phi) - f_1(\phi, X, Z) \rho_m + f_2(\phi, X, Z)] - \sum_{I=m,r} \int d^4x [e \rho_I (n_I) + {J_I}^\nu \partial_\nu \ell_{I}] \,.\label{mainaction_formula}
\end{equation}
In the above action formula, $f(T, \phi)$ represents a general function of the torsion scalar $T$ and the scalar field $\phi$, with $X:=-\frac{1}{2}\partial^{\mu}\phi\partial_{\mu}\phi$, serving as its kinetic term. The variable $Z$ is defined as $u^{\mu}_m \nabla_{\mu} \phi$, which combines the field derivative with the fluid's four-velocity related to DM, resulting in a scalar quantity.  In this context, we consider, the non-relativistic matter ($I = m$, which encompasses cold DM and baryons) and radiation ($I = r$). The energy density, denoted as $\rho_{I}$, varies based on the number density $n_I$. Moreover $\ell_{I}$ signifies a scalar field, while ${J_I}^\nu$ is the vector density.  The interaction between the scalar field and the cold DM fluid is incorporated through the functions $f_1$ and $f_2$, which are both dependent on $\phi$, $X$, and $Z$. The function $f_1$ facilitates the energy transfer, while $f_2$ is responsible for both the exchange of momentum and the scalar potential \cite{133_Amendola_2020,Rodriguez_Benites_2025}.

\subsection{Friedmann-Lemaître-Robertson-Walker Universe}\label{FLRWUniverse}

To probe cosmologicy in the scalar tensor gravity with a non-minimaly coupled scalar field we consider flat Friedmann-Lemaître-Robertson-Walker
(FLRW) spacetime, for which the tetrad field choice acan be written as,
\begin{equation}
    {e^A}_\mu = \text{diag}(1, a, a, a)\,,
\end{equation}
where a is the scale factor, and the metric equation can be presented in index form as,
\begin{equation}
    ds^2 = -dt^2 + a^2 \delta_{ij} dx^i dx^j \,.
\end{equation}
In this background setting, one can obtain the torsion scalar is $T = -6H^2$, where
$H = \dot{a}/a$ is the Hubble parameter, where dot represents the derivative with respect to the cosmic time $t$. The field equations in this setup can be obtained by varying the action Eq. \eqref{mainaction_formula} with respect to the tetrad
${e^A}_\mu$, and the scalar field $\phi$ as follow,
\begin{eqnarray}
     f - 2T f_{,T} - f_1 \rho_m + \rho_m f_{1,X} \phi^2 + \rho_m f_{1,Z} \dot{\phi} 
     + f_2 - f_{2,X} \dot{\phi}^2 - f_{2,Z} \dot{\phi} &=& \rho_m + \rho_r\,,\label{FE1}\\
     f - 2T f_{,T} - 4 \dot{H} f_{,T} - 4 H \dot{f}_{,T} + f_2 &=& -p_r \,.\label{FE2}\\
    (-\rho_m f_{1,ZZ} - \rho_m f_{1,X} - 2\dot{\phi} \rho_m f_{1,XZ} - 2X \rho_m f_{1,XX} + f_{2,ZZ} + f_{2,X} + 2\dot{\phi} f_{2,XZ} + 2X f_{2,XX}) \ddot{\phi} & \nonumber \\
    + 3H (\dot{\phi} f_{2,X} + f_{2,Z}) + \rho_m f_{1,\phi} - f_{2,\phi} \dot{\phi} f_{2,\phi Z} - \dot{\phi} \rho_m f_{1,\phi Z} \times 2X f_{2,\phi X} - 2X \rho_m f_{1,\phi X} - f_{,\phi} &=& 0\,.\label{KGE}
\end{eqnarray}
Here ($,$) denotes differentiation with respect to $\phi, Z, X$ or $T$. The definitions for functions $f, f_{1}, f_{2}$ can be presented as follow,

\begin{eqnarray}
    f &=& -\frac{T}{2\kappa^2}  - F(\phi)T^n \,,\\
    f_1 &=& \frac{1}{V_2(\phi)} - 1\,,\\
    f_2 &=& X \left[ 1 - \frac{1}{Y_1} + 2^{1-s/2} \beta \left( \frac{{Y_2}^s}{{Y_1}^{s/2}} \right) \,.\right]
\end{eqnarray}
From these equations one thing can be notted, we generalise $f$ in terms of general form of power law of torsion scalar $T$ \cite{Gonzalez-Espinoza:2020jss,Rodriguez_Benites_2025}. With this substitution, Eqs. \eqref{FE1}-\eqref{FE2} will takes the form as follows,
\begin{eqnarray}
    \frac{3H^2}{\kappa^2} &=& -F(\phi) T^n (2n-1)+ \tilde{\rho}_m+  \rho_r+ V_1 + (\beta + \frac{1}{2}) \dot{\phi}^2, \label{FFE1} \\
    -\frac{2\dot{H}}{\kappa^2} &=&\tilde{\rho}_m+ \frac{4}{3} \rho_r+(2\beta+1)\dot{\phi}^2+\dot{H}F(\phi)T^{n-1} [4n(2n-1)]+4 n H f_{\phi} \dot{\phi} T^{n-1}, \label{FFE2} 
    \end{eqnarray}
    The Klien Gordon equation in Eq. \eqref{KGE}  will take the form,
    \begin{eqnarray}
   (1+2\beta)\ddot{\phi} + 3H\dot{\phi}(2\beta+1) + T^{n} F_{,\phi} - \frac{\tilde{\rho}_m V_{2,\phi}}{V_2} + V_{1,\phi}=0. \label{eq:34}
\end{eqnarray}

In above equations we have $\tilde{\rho}_{m}=\frac{\rho_m}{V_{2}}$ ,which represents the effective matter energy density. The general Friedmann equations with cold DM, radiation and the DE can be presented as \cite{Gonzalez-Espinoza:2020jss},

\begin{eqnarray}
    \frac{3H^2}{\kappa^2} &=& \rho_{de} +\tilde{\rho}_{m} +\rho_{r}\,,\label{FriedmaanEq1}\\
    -\frac{2}{\kappa^2}\dot{H}&=&\rho_{de}+p_{de}+\tilde{\rho}_m+\frac{4}{3}\rho_r\label{FriedmaanEq2}\,.
\end{eqnarray}
On compairing Eqs. \eqref{FFE1},\eqref{FFE2} and \eqref{FriedmaanEq1},\eqref{FriedmaanEq2}, the equations for effective energy density and pressure can be obtained as, 

\begin{align}
    \rho_{de} &= (\beta + \frac{1}{2}) \dot{\phi}^2 + V_1 - F(\phi)T^n(2n-1) \\
    p_{de} &= (\beta + \frac{1}{2}) \dot{\phi}^2 - V_1 + 4n\dot{H} F(\phi)T^{n-1}(2n-1)+4nHF_{,\phi} \dot{\phi}T^{n-1}+F(\phi)T^{n}(2n-1) 
\end{align}
The effective EoS parameter can be obtained using following formula,
\begin{equation}
    w_{de} = \frac{p_{de}}{\rho_{de}}
\end{equation}
This teleparallel scalar tensor formalism also obey the conservation law and the fluid equations \cite{Rodriguez_Benites_2025,133_Amendola_2020} as follow,
\begin{eqnarray}
-\rho_m \dot{f}_1&=&\dot{\rho}_{de} + 3H (\rho_{de} + p_{de}) \\
+\rho_m \dot{f}_1&=&\dot{\tilde{\rho}}_m + 3H \tilde{\rho}_m \\
0&=& \dot{\rho}_r + 4H \rho_r 
\end{eqnarray}
Moreover the equations for deceleration parameter and EoS ($\omega_{tot}$) can be calculated using,
\begin{align}
     w_{tot} = \frac{p_{de} + p_r}{\rho_{de} + \tilde{\rho}_m + \rho_r}=\frac{1}{3}(-1+2q) \quad
\end{align}

The equation of density parameters for effective radiation, matter and DE can be written as,

\begin{align}
  \Omega_m \equiv \frac{\kappa^2 \tilde{\rho}_m}{3H^2} \quad
    \Omega_r \equiv \frac{\kappa^2 \rho_r}{3H^2} \quad  \Omega_{de} \equiv \frac{\kappa^2 \rho_{de}}{3H^2}
\end{align}
which satisfy the constrain equation as given below,
\begin{equation}
    \Omega_{de} + \Omega_m + \Omega_r = 1
\end{equation}
\subsection{Dynamical Analysis and Cosmological Parameters in a Power Law Model}\label{dynamical_system_analysis}
In this section we will be discussing detail analysis of autonomous dynamical system framed using evolution Eqs. \eqref{FFE1},\eqref{FFE2} and the Klein Gordon equation presented in \eqref{KGE} of this scalar tensor gravity model. The dynamical variables in this case are \cite{Rodriguez_Benites_2025,Gonzalez-Espinoza:2020jss}. 
\begin{align}
x &= \frac{\kappa \dot{\phi}}{\sqrt{6} H}, & y &= \frac{\kappa \sqrt{V_1}}{\sqrt{3} H}, & u &= -\frac{\kappa^2 F(\phi) T^n (2n-1)}{3H^2}, \\
\Omega_m &= \frac{\kappa^2 \rho_m}{V_2 3 H^2}, & \lambda &= \frac{V_{1,\phi}}{\kappa V_1}, & \sigma &= -\frac{F_{,\phi}}{\kappa F}, \\
Q &= -\frac{V_{2,\phi}}{\kappa V_2}, & \Theta &= \frac{F F_{,\phi\phi}}{(F_{,\phi})^2}, & \Gamma_1 &= \frac{V_1 V_{1,\phi\phi}}{(V_{1,\phi})^2}, \\
\Gamma_2 &= \frac{V_2 V_{2,\phi\phi}}{(V_{2,\phi})^2}, & \varrho &= \frac{\kappa \sqrt{\rho_r}}{\sqrt{3} H}. \label{dynamical_variables}
\end{align}

These dynamical parameters obey the constrain equation can be written as follow,
\begin{equation}
1=u+\Omega_m + \varrho^2 + y^2+(1 + 2\beta) x^2 
\end{equation}
We have chosen exponential potential as \cite{Rodriguez_Benites_2025,Gonzalez-Espinoza:2020jss}, 

The autonomous dynamical system can be framed by differentiating dynamical variables in Eq. \eqref{dynamical_variables} with respect to e-folding number $N=log(a)$ and can be obtained as follow,
\begin{align}
    \frac{dx}{dN}=&\frac{\sqrt{6} x^2}{2} \left(Q-\frac{2 n \sigma  u}{(2 n-1) (n u-1)}\right)+\frac{\sqrt{6} \left((2 n-1) \left(Q \left(\rho ^2+u+y^2-1\right)+\lambda  y^2\right)-\sigma  u\right)}{2(2 \beta +1) (2 n-1)}\nonumber\\
    &-\frac{3 (2 \beta +1) x^3}{2(n u-1)}+\frac{x \left((3-6 n) u-\rho ^2+3 y^2+3\right)}{2(n u-1)}\,,\nonumber\\
    \frac{dy}{dN}=&-\frac{\sqrt{6}}{2} \lambda  x y+\frac{3y \left(\frac{2 \sqrt{\frac{2}{3}} n \sigma  u x}{1-2 n}-\frac{\rho ^2}{3}+u-\left((2 \beta +1) x^2\right)+y^2-1\right)}{2(n u-1)}\,,\nonumber\\
    \frac{du}{dN}&=-\left(\sqrt{6}u\right) \sigma  x u-\frac{3 n u\left(\frac{2 \sqrt{\frac{2}{3}} n \sigma  u x}{1-2 n}-\frac{\rho ^2}{3}+u-\left((2 \beta +1) x^2\right)+y^2-1\right)}{n u-1}\nonumber\\
    &+\frac{3 u \left(\frac{2 \sqrt{\frac{2}{3}} n \sigma  u x}{1-2 n}-\frac{\rho ^2}{3}+u-\left((2 \beta +1) x^2\right)+y^2-1\right)}{n u-1}\,,\nonumber\\
    \frac{d\rho}{dN}&=-2 \rho+\frac{3 \rho  \left(\frac{2 \sqrt{\frac{2}{3}} n \sigma  u x}{1-2 n}-\frac{\rho ^2}{3}+u-\left((2 \beta +1) x^2\right)+y^2-1\right)}{2 (n u-1)}\,.\label{dynamical_systme_equations}
\end{align}
From Eqs. \eqref{FE1} and \eqref{FE2}, and using the dynamical variables defined in Eq. \eqref{dynamical_variables} one can easily obtain,
\begin{equation}
    \frac{\dot{H}}{H^2}=-\frac{3 \left(\frac{2 \sqrt{\frac{2}{3}} n \sigma  u x}{1-2 n}-\frac{\rho ^2}{3}+u-\left((2 \beta +1) x^2\right)+y^2-1\right)}{2 (n u-1)}
\end{equation}
This equation is essential because it allows for the direct calculation of the deceleration parameter ($q$) and the values of the equation of state ($\omega_{tot}$) through its use. Here the expressions for deceleration parameter ($q$, $\omega_{de}$ and $\omega_{tot}$) can be calculated as,
\begin{align}
    q&=-1+\frac{3 \left(\frac{2 \sqrt{\frac{2}{3}} n \sigma  u x}{1-2 n}-\frac{\rho ^2}{3}+u-\left((2 \beta +1) x^2\right)+y^2-1\right)}{2 (n u-1)}\,,\nonumber\\
    \omega_{tot}&=-1+\frac{\frac{2 \sqrt{\frac{2}{3}} n \sigma  u x}{1-2 n}-\frac{\rho ^2}{3}+u-\left((2 \beta +1) x^2\right)+y^2-1}{n u-1}\,,\nonumber\\
    \omega_{de}&=\frac{u \left(n \left(-2 n \left(\rho ^2+3\right)+\rho ^2-2 \sqrt{6} \sigma  x+9\right)-3\right)+3 (2 n-1) \left(y^2-(2 \beta +1) x^2\right)}{3 (2 n-1) (n u-1) \left(u+(2 \beta +1) x^2+y^2\right)}\,.
\end{align}
Moreover in this case we have expressions for standard density parameter for effective matter radiation and the  DE can be written as follow,
\begin{align}
    \Omega_{m}&=-\rho ^2-u-\left((2 \beta +1) x^2\right)-y^2+1\,,\nonumber\\
    \Omega_{de}&=u+(2 \beta +1) x^2+y^2\,,\nonumber\\
    \Omega_{r}&=\rho^2\,.
\end{align}
\subsection{Analysis of Critical Points and Their Stability}\label{critical_point_analyais}
In this study we have obtained the critical points by considering $\frac{dx}{dN}=\frac{dy}{dN}=\frac{du}{dN}=\frac{d\rho}{dN}=0$. In this study we are confined ourself to the exponential potentials which make the further analysis quit simpler the forms of coupling potentials and scalar functions that we have cohosen are $V_{1}=e^{-\kappa \lambda \phi}, V_{2}=e^{-\kappa Q \phi}, F=e^{-\kappa \sigma \phi}$. There are studies related to solutions leading to explain late time cosmic acceleration and the scaling solutions these functions have been analysed\cite{Gonzalez-Espinoza:2020jss,copelandLiddle,Bahamonde_2019,Zlatev_1999}. We have obtained the eigenvalues of the Jacobian matrix for the above systme of equations presented in Eq. \eqref{dynamical_systme_equations}. Depending upon sign of the eigenvalues we can obtain the stability. All types of stability are covered further as The eigenvalues of the Jacobian matrix specifically determine the stability criteria for critical points \cite{Copeland:2006wr,G.Otalora2013JCAPDSA}: (i) Stable node: If all eigenvalues exhibit negative value at the relevant critical point. (ii) Unstable node: If all eigenvalues exhibit positive values. (iii) Saddle point: If one or two of the three eigenvalues are positive while the remaining ones are negative. (iv) Stable spiral: If the determinant of the Jacobian matrix is negative and the real parts of the eigenvalues are negative. A critical point acts as an attractor in the first and fourth scenarios, but not in the second and third. The critical points are presented in the below Table \ref{critical_points}, with the eigenvalues with the values of standard density parameters in Table \ref{density_parameters}. The detail analysis of these critical points can be presented as follow,
The detailed analysis of the critical points is presented as follows,
\begin{itemize}
\item{ Critical Point \textbf{$A_R$}}:
This critical point is a standard radiation dominated critical point with $\Omega_{r}=1$. Here at this critical point the values of EoS parameters are $\omega_{tot}=\frac{1}{3}, \omega_{de}=1$. The deceleration parameter $q=1$, at this critical point and can be observed from Table \ref{critical_points}. The eigenvalues at this critical points are $\{\gamma_{1}=-1,\gamma_{2}=1,\gamma_{3}=2,\gamma_{4}=4-4 n\}$. Since presence of both the positive and negative eigenvalues implies this critical point to be saddle critical point. Moreover this critical poin is in agreement to the study made in \cite{Rodriguez_Benites_2025}. \\
\end{itemize}

\begin{table}[H]
    \centering 
    \begin{tabular}{|c |c |c |c| c| c|} 
    \hline\hline 
    \parbox[c][0.9cm]{1.3cm}{{Name}
    }& $ \{ x_{c}, \, y_{c}, \, u_{c}, \, \rho_{c} \} $ &  {$\omega_{tot}$}& {$\omega_{de}$}& \parbox[c][0.9cm]{0.9cm}{$q$} \\ [0.5ex] 
    \hline\hline 
    \parbox[c][1.3cm]{1.3cm}{$A_{R}$ } &$\{ 0, 0, 0, 1 \}$ &  $\frac{1}{3}$& $1$ & $1$\\
    \hline
    \parbox[c][1.3cm]{1.3cm}{$B$ } & $\{0, \,  \frac{\sqrt{\sigma }}{\sqrt{-\lambda +2 \lambda  n+\sigma }}, \, \frac{\lambda  (2 n-1)}{-\lambda +2 \lambda  n+\sigma } ,  \, 0\}$ &  $-1$&  $-1$ & $-1$\\
    \hline
    \parbox[c][1.3cm]{1.3cm}{$C_{R}$} & $\{-\frac{1}{\sqrt{6} Q}, \,  0, \, 0 ,  \, \frac{\sqrt{-\frac{2 \beta }{Q}+2 Q-\frac{1}{Q}}}{\sqrt{2} \sqrt{Q}}\}$ &  $\frac{1}{3}$&  $1$ & $1$\\
    \hline
   \parbox[c][1.3cm]{1.3cm}{$D_{M}$ } &  $\{-\frac{\sqrt{\frac{2}{3}} Q}{2 \beta +1}, \, 0, \, 0, \, 0 \}$ &  $\frac{2 Q^2}{6 \beta +3}$& $1$ & $\frac{1}{2}+\frac{Q^2}{2 \beta +1}$\\
   \hline
   \parbox[c][1.3cm]{1.3cm}{$E^{\pm}$} &   $\left\{\pm\frac{1}{\sqrt{2 \beta +1}},0,0,0\right\}$ &  $1$& $1$ & $2$ \\
   \hline
   \parbox[c][1.3cm]{1.3cm}{$F_{R}$} & $\{\frac{2 \sqrt{\frac{2}{3}}}{\lambda },\frac{2 \sqrt{2 \beta +1}}{\sqrt{3} \lambda },0, \frac{\sqrt{-8 \beta +\lambda ^2-4}}{\lambda }\}$ &  $\frac{1}{3}$&  $\frac{1}{3}$ & $1$\\
   \hline
\parbox[c][1.3cm]{1.3cm}{$G$} & 
$\left\{ \frac{\lambda }{\sqrt{6} (2 \beta +1)}, 
\frac{\sqrt{\frac{\left(12 \beta -\lambda ^2+6\right) (\lambda +Q)}{2 \beta +1}}}{\sqrt{6} \sqrt{\lambda +Q}}, 
0,0 \right\}$ & 
$-1+\frac{\lambda ^2}{6 \beta +3}$ & 
$-1+\frac{\lambda ^2}{6 \beta +3}$ & 
$-1+\frac{\lambda ^2}{4 \beta +2}$ \\
 \hline
\parbox[c][1.3cm]{1.3cm}{$H$} & 
$\left\{ \frac{\sqrt{\frac{3}{2}}}{\lambda +Q}, 
\frac{\sqrt{6 \beta +2 Q^2+2 \lambda  Q+3}}{\sqrt{\lambda +Q} \sqrt{2 \lambda +2 Q}}, 
0,0 \right\}$ & 
$-\frac{Q}{\lambda +Q}$ & 
$-\frac{Q (\lambda +Q)}{6 \beta +Q (\lambda +Q)+3}$ & 
$\frac{\lambda -2 Q}{2 (\lambda +Q)}$ \\
\hline
\parbox[c][1.5cm]{1.5cm}{$I_{M} (n=1)$} & 
$\left\{ 0, 0, \frac{Q}{Q-\sigma}, 0 \right\}$ & 
$0$ & 
$0$ & 
$\frac{1}{2}$ \\
 \hline
\parbox[c][1.3cm]{1.3cm}{$J_{R}$} & 
$\left\{ -\frac{2 \sqrt{\frac{2}{3}} (n-1)}{\sigma }, 0,\frac{4 (2 \beta +1) \left(2 n^2-3 n+1\right)}{3 \sigma ^2}, \frac{\sqrt{\sigma ^2-\frac{4}{3} (2 \beta +1) (n-1) (4 n-3)}}{\sigma }\right\}$ & 
$\frac{1}{3}$ & 
$\frac{1}{3}$ & 
$1$ \\
\hline
 \end{tabular}
\caption{Critical points for the autonomous dynamical system in Eq. \eqref{dynamical_systme_equations} for power law model.}
\label{critical_points}
\end{table}

\begin{table}[H]
    \centering 
    \begin{tabular}{|c |c |c |c|} 
    \hline\hline 
    \parbox[c][0.9cm]{1.3cm}{{Name}
    }& {$\Omega_{r}$}& {$\Omega_{m}$}& \parbox[c][0.9cm]{0.9cm}{$\Omega_{de}$} \\ [0.5ex] 
    \hline\hline 
    \parbox[c][1.3cm]{1.3cm}{$A_{R}$ } & $1$& $0$ & $0$\\
    \hline
    \parbox[c][1.3cm]{1.3cm}{$B$ } & $0$&  $0$ & $1$\\
    \hline
    \parbox[c][1.3cm]{1.3cm}{$C_{R}$ } & $1-\frac{\beta +\frac{1}{2}}{Q^2}$&  $\frac{2 \beta +1}{3 Q^2}$ & $\frac{2 \beta +1}{6 Q^2}$ \\
    \hline
   \parbox[c][1.3cm]{1.3cm}{$D_{M}$ } & $0$& $1-\frac{2 Q^2}{6 \beta +3}$ & $\frac{2 Q^2}{6 \beta +3}$\\
   \hline
   \parbox[c][1.3cm]{1.3cm}{$E^{\pm}$} & $0$& $0$ & $1$\\
   \hline
   \parbox[c][1.3cm]{1.3cm}{$F_{R}$} & $1-\frac{8 \beta +4}{\lambda ^2}$&  $0$ & $\frac{8 \beta +4}{\lambda ^2}$ \\
   \hline
 \parbox[c][1.3cm]{1.3cm}{$G$} & $0$ & $0$ & $1$\\
 \hline
 \parbox[c][1.3cm]{1.3cm}{$H$} & $0$ & $\frac{-6 \beta +\lambda  (\lambda +Q)-3}{(\lambda +Q)^2}$ & $\frac{6 \beta +Q (\lambda +Q)+3}{(\lambda +Q)^2}$ \\
 \hline
  \parbox[c][1.3cm]{1.3cm}{$I_{M}$} & $0$ & $1-\frac{Q}{Q-\sigma }$ & $\frac{Q}{Q-\sigma }$ \\
 \hline
  \parbox[c][1.3cm]{1.3cm}{$J_{R}$} & $1-\frac{4 (2 \beta +1) (n-1) (4 n-3)}{3 \sigma ^2}$ & $0$ & $\frac{4 (2 \beta +1) (n-1) (4 n-3)}{3 \sigma ^2}$ \\
  \hline
 \end{tabular}
\caption{Values of Standard Density Parameters for power law model.}
\label{density_parameters}
\end{table}
\begin{itemize}
\item{\textbf{Critical Points} $B$}: The critical point $B$ is the de-Sitter solution. This critical points attains the value of deceleration parameter and EoS parameter to be $-1$. The eigenvalues at this critical points are as follow
$\left\{\gamma_{1}= -2, \right. 
\quad \gamma_{2}=\text{Root} \left[P, 1 \right],
\quad \gamma_{3}=\text{Root} \left[P, 2 \right], 
\quad \gamma_{4}=\text{Root} \left[ P, 3 \right] \left. \right\}$ 
Here$\gamma_{2},\gamma_{3},\gamma_{4}$ are the second, third and fourth root of the function P which takes the form as $P=\left[ \frac{72 \lambda^8 \sigma \sqrt{-\lambda + 2n \lambda + \sigma} - 1512 n \lambda^8 \sigma \sqrt{-\lambda + 2n \lambda + \sigma} }{2 (-1 + 2n) (1 + 2\beta) (\lambda - 3n \lambda + 2n^2 \lambda - \sigma) (-\lambda + 2n \lambda + \sigma)^{3/2}}\right]$   This critical point show stability at
asymptotically Stable if $\sigma > 0$ and $n >\frac{1}{2}$, or, $\sigma < 0$ and $n < \frac{1}{2}$
Unstable if $\sigma > 0$ and $n < \frac{1}{2}$, or $\sigma < 0$ and $n > \frac{1}{2}$. This critical point is the standard DE critical point, $\Omega_{de}=1$ and the values of the other two (radiation and matter) standard density parameters are $0$.

\item{\textbf{Critical Point $C_{R}:$} This critical point describes a non-standard radiation-dominated critical point with a value of $\Omega_{r}=1-\frac{\beta +\frac{1}{2}}{Q^2}$ some contribution of the standard matter density parameter $\frac{\beta +\frac{1}{2}}{Q^2}$. This is the radiation dominated scaling solution and as discussed in \cite{Rodriguez_Benites_2025}, this critical point may describe early time Big Bang Nucleosynthesis (BBN) constraints of $\Omega^{r}_{de}\le 0.045$ \cite{Ferreira:1998,Bean:2001}
The eigenvalues at this critical point are presented as follows:
$\Big\{\gamma_{1}=-\frac{1}{2}-\frac{\sqrt{(2 \beta +1)^2 \left(-(1-2 n)^2\right) Q^2 \left(-4 \beta +3 Q^2-2\right)}}{2 (2 \beta +1) (2 n-1) Q^2},\gamma_{2}=\frac{1}{2} \left(\frac{\sqrt{(2 \beta +1)^2 \left(-(1-2 n)^2\right) Q^2 \left(-4 \beta +3 Q^2-2\right)}}{(2 \beta +1) (2 n-1) Q^2}-1\right),\gamma_{3}=\frac{\lambda }{2 Q}+2,\\ \gamma_{4}=-4 n+\frac{\sigma }{Q}+4\Big\}$}.\\ This point shows stable behaviour at\\
$n\in \mathbb{R}\land n-\frac{1}{2}\neq 0\land \left(Q<0\land \sigma >4 n Q-4 Q\land \frac{1}{4} \left(3 Q^2-2\right)\leq \beta <\frac{1}{2} \left(2 Q^2-1\right)\land \lambda >-4 Q\right)$ or $n\in \mathbb{R}\land n-\frac{1}{2}\neq 0\land \left(Q>0\land \sigma <4 n Q-4 Q\land \frac{1}{4} \left(3 Q^2-2\right)\leq \beta <\frac{1}{2} \left(2 Q^2-1\right)\land \lambda <-4 Q\right)$. At this critical points, values of $\omega_{tot}=\frac{1}{3}, \omega_{de}=1, q=1$ hence does not describe the late time accelerating phenomena. Moreover this critical point describe standard radiation dominated era at $\beta=-\frac{1}{2}$.

\item{\textbf{Critical Point $D_{M}:$}} This is the non-standard matter dominated critical point. At this critical point the standard density for matter and DE contributes. This critical point is the matter dominated scaling solution with $\Omega_{de}=\frac{2 Q^2}{6 \beta +3}$. This critical point will describe the standard matter dominated solution at $Q=0$, at which $\Omega_{m}=1, \Omega_{de}=0$. The eigenvalues at the Jacobian matrix can be calculated as, $\Big\{\gamma_{1}=-\frac{2 \beta -2 Q^2+1}{2 (2 \beta +1)}, \gamma_{2}=-\frac{6 \beta -2 Q^2+3}{2 (2 \beta +1)},\gamma_{3}=\frac{6 \beta +2 Q^2+2 \lambda  Q+3}{2 (2 \beta +1)},\gamma_{4}=-\frac{-6 \beta +6 \beta  n+2 n Q^2+3 n-2 Q^2-2 Q \sigma -3}{2 \beta +1}\Big\}$ and show stability at $n<1\land \bigl(\sigma <-\sqrt{8 n^2-16 n+8}\land \bigl(Q<0\land \frac{-2 n Q^2-3 n+2 Q^2+2 Q \sigma +3}{6 n-6}<\beta <-\frac{1}{2}\land \lambda <\frac{-6 \beta -2 Q^2-3}{2 Q}\bigl)\bigl)$. As discussed in \cite{Rodriguez_Benites_2025}, this critical point bound to describe CMB constrain  $\Omega^{m}_{de}<0.02$ at redshift $z\approx 50$ \cite{2016_Planck}.

\item{\textbf{Critical Point $E^{\pm}:$}}
This critical point represent DE-dominated era with $\Omega_{de}=1$, but is not describing current accelerating phase of the Universe evolution, but it describe the stiff matter dominated era with $\omega_{de}=\omega_{tot}=1$. The eigenvalues at this critical points are
$\Big\{\gamma_{1}=1,\gamma_{2}=\frac{3 \sqrt{2 \beta +1}\pm\sqrt{6} Q}{\sqrt{2 \beta +1}},\gamma_{3}=\frac{6 \sqrt{2 \beta +1}\pm\sqrt{6} \lambda }{2 \sqrt{2 \beta +1}},\gamma_{4}=-\frac{-6 \sqrt{2 \beta +1}+6 \sqrt{2 \beta +1} n\pm\sqrt{6} \sigma }{\sqrt{2 \beta +1}}\Big\}$. This critical point shows unstable behaviour at $\sigma \in \mathbb{R}\land \bigl(n<\frac{1}{2}\land \lambda >2 \sigma \land -\frac{1}{2}<\beta <\frac{6 \lambda ^2-24 \lambda  \sigma -144 n^2+144 n+24 \sigma ^2-36}{288 n^2-288 n+72}\land Q<-\frac{\lambda }{2}\bigl)$, moreover is saddle point within the range $\sigma \in \mathbb{R}\land \left(n<\frac{1}{2}\land \left(\lambda \leq 2 \sigma \land \beta >-\frac{1}{2}\land Q<-\frac{\lambda }{2}\right)\right)$. Presence of positive eigenvalue $1$ at this critical point makes this critical point saddle and unstable.

\item{\textbf{Critical Point $F_{R}:$}}
This critical point is a non-standard radiation dominated era. Here both standard density parameter for matter and DE will contribute. The value of $\Omega_{de}=\frac{8\beta+4}{\lambda^2}$, and hence is a rediation dominated scaling solution. This critical point should describe the BBN constrain from the early Universe on the value of standard DE density parameter $\Omega^{r}_{de}<0.045$ \cite{Ferreira:1998,Bean:2001}. This critical point show stability at $\bigl(n\in \mathbb{R}\land n-\frac{1}{2}\neq 0\land \lambda <0\land Q>-\frac{\lambda }{4}\land \frac{1}{128} \left(15 \lambda ^2-64\right)\leq \beta <\frac{1}{8} \left(\lambda ^2-4\right)\land \sigma <\lambda -\lambda  n\bigl)\lor \bigl(n\in \mathbb{R}\land n-\frac{1}{2}\neq 0\land \lambda >0\land Q<-\frac{\lambda }{4}\land \frac{1}{128} \left(15 \lambda ^2-64\right)\leq \beta <\frac{1}{8} \left(\lambda ^2-4\right)\land \sigma >\lambda -\lambda  n\bigl)$, with the eigenvalues,

$\Big\{\gamma_{1}=\frac{4 Q}{\lambda }+1,\gamma_{2}=-\frac{\sqrt{(2 \beta +1)^2 \lambda ^2 (1-2 n)^2 \left(128 \beta -15 \lambda ^2+64\right)}}{2 (2 \beta +1) \lambda ^2 (2 n-1)}-\frac{1}{2}, \gamma_{3}=\frac{1}{2} \left(\frac{\sqrt{(2 \beta +1)^2 \lambda ^2 (1-2 n)^2 \left(128 \beta -15 \lambda ^2+64\right)}}{(2 \beta +1) \lambda ^2 (2 n-1)}-1\right),\gamma_{4}=-\frac{4 \sigma }{\lambda }-4 n+4\Big\}$. This critical point represent the standard radiation dominated era at $\beta=\frac{-1}{2}$, where $\Omega_{r}=1,\Omega_{de}=0.$

\item{$\textbf{Critical Point G}:$} This is an DE dominated critical point with $\Omega_{de}=1$. The value of $\omega_{tot}=-1+\frac{\lambda ^2}{6 \beta +3}$ This critical point sucessfully explain the current accelerating phase of the Universe evolution within the range $\lambda \in \mathbb{R}\land \left(\beta <-\frac{1}{2}\lor \beta >\frac{1}{4} \left(\lambda ^2-2\right)\right)$, where the value of $\omega_{tot}$ is less than $\frac{-1}{3}.$ The eigenvalues at this critical point will  $\left\{ \gamma_{1}=\frac{\lambda ^2}{4 \beta +2}-3,\gamma_{2}=\frac{\lambda ^2}{4 \beta +2}-2,\gamma_{3}=\frac{-6 \beta +\lambda  (\lambda +Q)-3}{2 \beta +1},\gamma_{4}=-\frac{\lambda  (\lambda  (n-1)+\sigma )}{2 \beta +1}\right\}$ and show stability within the range $\sigma \in \mathbb{R}\land \bigl(\bigl(\lambda <0\land \bigl(\left(\beta <-\frac{1}{2}\land Q<\frac{6 \beta -\lambda ^2+3}{\lambda }\land n<\frac{\lambda -\sigma }{\lambda }\right)\lor \bigl(\beta >\frac{1}{8} \left(\lambda ^2-4\right)\land Q>\frac{6 \beta -\lambda ^2+3}{\lambda }\land n>\frac{\lambda -\sigma }{\lambda }\bigl)\bigl)\bigl)\lor \bigl(\lambda >0\land \left(\bigl(\beta <-\frac{1}{2}\land Q>\frac{6 \beta -\lambda ^2+3}{\lambda }\land n<\frac{\lambda -\sigma }{\lambda }\right)\lor \bigl(\beta >\frac{1}{8} \left(\lambda ^2-4\right)\land Q<\frac{6 \beta -\lambda ^2+3}{\lambda }\land n>\frac{\lambda -\sigma }{\lambda }\bigl)\bigl)\bigl)\bigl)$. This critical point describe both quentessence and the phantom region which will be depending on the value of $\lambda$, in both the cases critical point will describe the current phase of cosmic acceleration. 

\item{$\textbf{Critical Point H:}$}
This critical point represents matter-dominated scaling solution with $\Omega_{de}= \frac{6 \beta +Q (\lambda +Q)+3}{(\lambda +Q)^2}, \Omega_{m}=\frac{-6 \beta +\lambda  (\lambda +Q)-3}{(\lambda +Q)^2}$. This critical point must adhere to satisfy $\Omega^{m}_{de}< 0.02$  at a redshift of $z \approx 50$, based on CMB measurements \cite{2016_Planck}. This critical point will describe standard matter dominated era at $\beta =\frac{1}{6} \left(-Q^2-\lambda  Q-3\right)\land \lambda +Q\neq 0$, where $\Omega_{m}=1$. Theegienvalues at this critical points are, $\left\{-\frac{\lambda +4 Q}{2 (\lambda +Q)},-\frac{\gamma +3 (\lambda +Q) (\lambda +2 Q)}{4 (\lambda +Q)^2},\frac{\gamma -3 (\lambda +Q) (\lambda +2 Q)}{4 (\lambda +Q)^2},-\frac{3 (\lambda  (n-1)+\sigma )}{\lambda +Q}\right\}$ and will show stable behaviour within the range $\lambda =0\land \bigl(\bigl(n\in \mathbb{R}\land n-\frac{1}{2}\neq 0\land Q<0\land \frac{1}{96} \left(-20 Q^2-48\right)-\frac{5 \sqrt{Q^4}}{24}\leq \beta <\frac{1}{6} \left(-2 Q^2-3\right)\land \sigma <0\bigl)\lor \bigl(n\in \mathbb{R}\land n-\frac{1}{2}\neq 0\land Q>0\land \frac{1}{96} \left(-20 Q^2-48\right)-\frac{5 \sqrt{Q^4}}{24}\leq \beta <\frac{1}{6} \left(-2 Q^2-3\right)\land \sigma >0\bigl)\bigl)$\\

here, $\gamma =\frac{\sqrt{3} \sqrt{(2 \beta +1) (1-2 n)^2 (\lambda +Q)^2 \left(3 (2 \beta +1) \left(48 \beta -7 \lambda ^2+24\right)-16 \lambda  Q^3+4 Q^2 \left(30 \beta -8 \lambda ^2+15\right)+4 \lambda  Q \left(18 \beta -4 \lambda ^2+9\right)\right)}}{(2 \beta +1) (2 n-1)}$

\item{$\textbf{Critical Point $I_M$:}$}
This critical point will exist at particular $n=1$ case. This is a matter dominated scaling solution with $\omega_{de}=\frac{Q}{Q-\sigma }$. The value of $\omega_{de}=\omega_{tot}=0$ and $q=\frac{1}{2}$. This critical point will describe the standard matter dominated criticl point at $Q=0.$  As one can clearlly note these values are depend on both
the nonminimal coupling to
 gravity $\sigma$ and the energy coupling $Q$.  The eigenvalues at this critical point are
$\Big\{\gamma_{1}=\frac{3}{2},\gamma_{2}=-\frac{1}{2},\gamma_{3}=\frac{3 (2 \beta +1) \sigma ^3-\sqrt{3} \mu -3 (2 \beta +1) Q \sigma ^2}{4 (2 \beta +1) \sigma ^2 (Q-\sigma )},\gamma_{4}=\frac{3 (2 \beta +1) \sigma ^3+\sqrt{3} \mu -3 (2 \beta +1) Q \sigma ^2}{4 (2 \beta +1) \sigma ^2 (Q-\sigma )}\Big\}$
where $\mu =\sqrt{(2 \beta +1) \sigma  (Q-\sigma ) \left(-3 (2 \beta +1) \sigma ^4-16 Q^2 \sigma ^4+Q \sigma ^3 \left(3 (2 \beta +1)+16 \sigma ^2\right)\right)}$
The presence of positive $\frac{3}{2} $ and negative $-\frac{1}{2}$ eigenvalues imply this critical point as a saddle critical point.
\item{$\textbf{Critical Point $J_R$}:$} This is an additional radiation-dominated scaling solution with $\Omega_{de}=\frac{4 (2 \beta +1) (n-1) (4 n-3)}{3 \sigma ^2}$. Moreover this critical point can describe the BBN constrain on standard density parameter on DE $\Omega^{r}_{de}<0.045$ \cite{Ferreira:1998,Bean:2001}. This critical point will describe the standard density parameter at the values of parameter $n=1,\frac{3}{4}$ and $\beta=\frac{-1}{2}$, where $\Omega_{r}=1$. The eigenvalues at this critical point are
$\left\{ 2 + \frac{2(n-1)\lambda}{\sigma}, \frac{4Q(1-n) + \sigma}{\sigma}, \frac{A - \sqrt{B}}{D}, \frac{A + \sqrt{B}}{D}\right\}$\\
Where
\begin{align*}
A &= (2n-1)(1+2\beta)\sigma\left[-4(n-1)n(2n-1)(1+2\beta) + 3\sigma^2\right], \\
D &= 2(2n-1)(1+2\beta)\sigma\left[4(n-1)n(2n-1)(1+2\beta) - 3\sigma^2\right], \\
B &= (1-2n)^2(1+2\beta)^2\sigma^2 \times \\
   &\quad \left[16(n-1)^2n(2n-1)(48 + 66n^2 - 113n)(1+2\beta)^2 \right. \\
   &\quad \left. - 24(n-1)(8n^3 + 5n^2 - 49n + 24)(1+2\beta)\sigma^2 \right. \\
   &\quad \left. + 9(16n-15)\sigma^4\right]\,.
\end{align*}
and show stability within the parametric range $n\in \mathbb{R}\&\&n-\frac{1}{2}\neq 0$ and $\lambda =0\land \sigma <0\land \frac{1}{96} \left(-20 Q^2-48\right)-\frac{5 \sqrt{Q^4}}{24}\leq \beta <\frac{1}{6} \left(-2 Q^2-3\right)$. This is an critical point which occur in addition to all the critical points studied in \cite{Rodriguez_Benites_2025}, others critical point are occuring with the inclusion of parameter $n$ in the study. 
\end{itemize}
\subsection{Numerical Results}\label{Numerical_results}
\begin{figure}[H]
\centering
\begin{subfigure}[h]{0.3\textwidth}
\centering
\includegraphics[width=78mm]{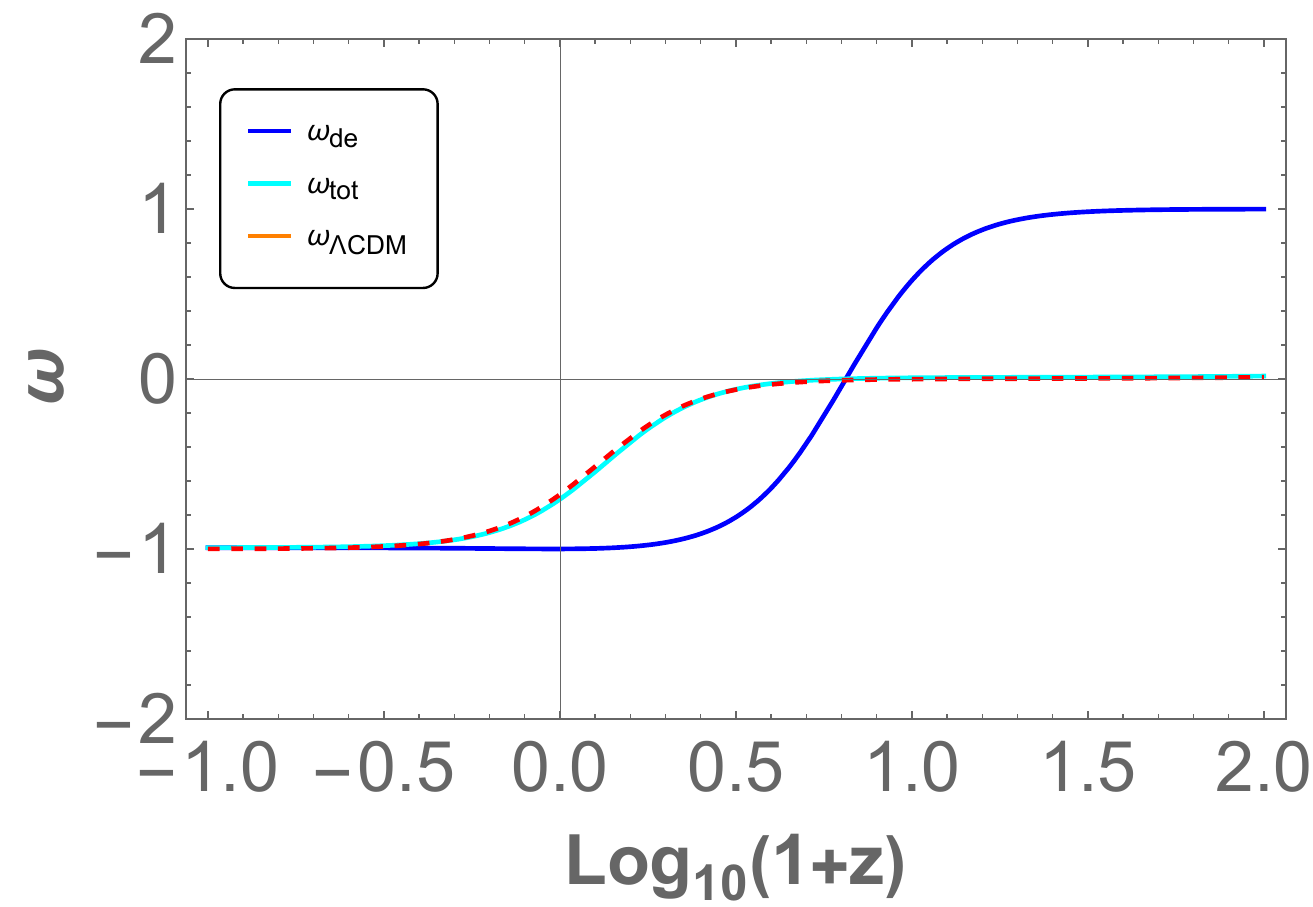}
\caption{EoS parameters}\label{Eosmfig}
\end{subfigure}
\hspace{1.9cm}
\begin{subfigure}[h]{0.5\textwidth}
\centering
\includegraphics[width=78mm]{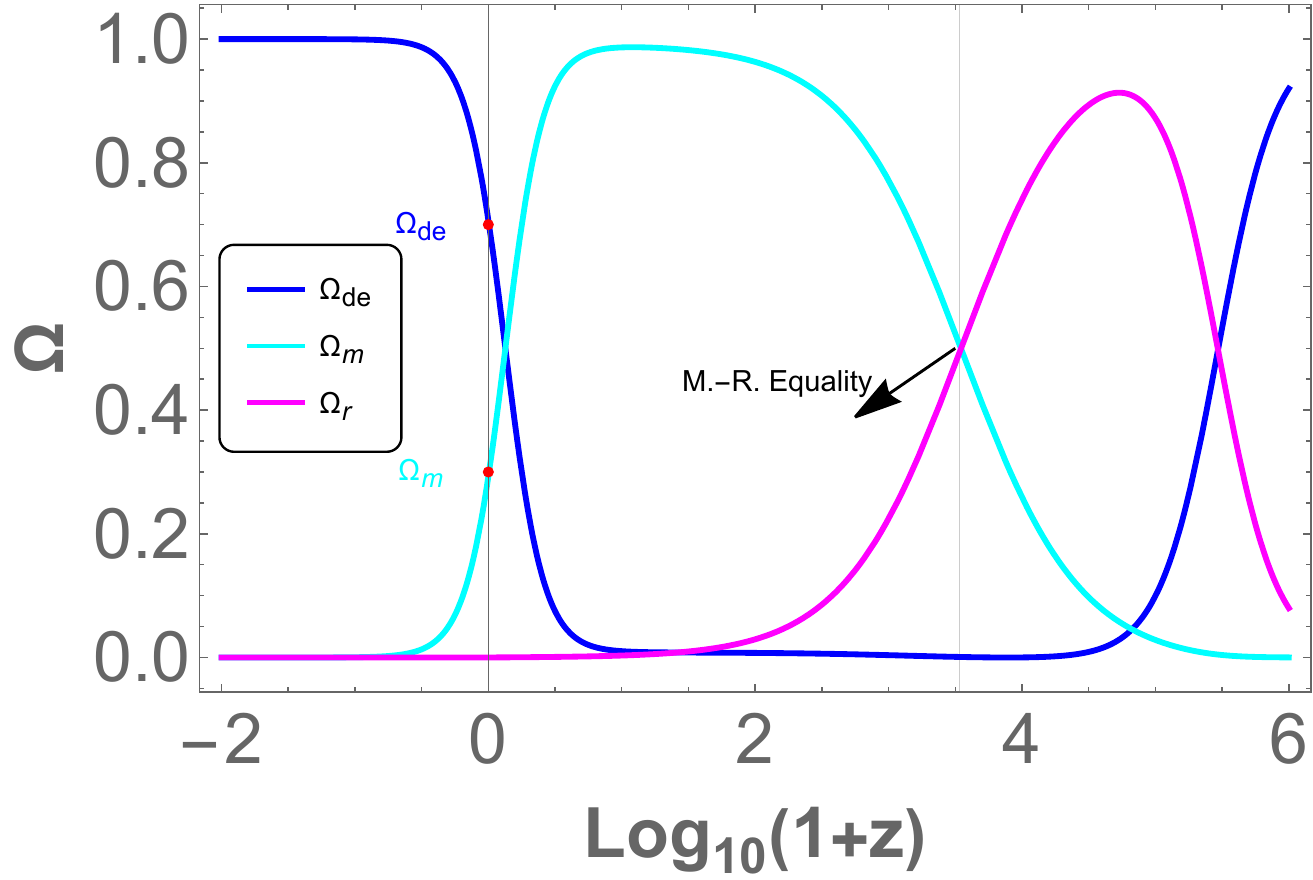}
\caption{Standard density parameters}\label{densityparametersfig}
\end{subfigure}
\caption{The initial conditions are $x_C=10^{-4} ,\,y_C=10^{-6} ,\,u_C=10^{-8} ,\,\rho_C=0.883 ,\, n=2, \lambda=0.8, \sigma=12.2, \beta=14.05, Q=0.6$. }\label{Eosdensitym1}
\end{figure}

\begin{figure}[H]
\centering
\begin{subfigure}[h]{0.35\textwidth}
\centering
\includegraphics[width=78mm]{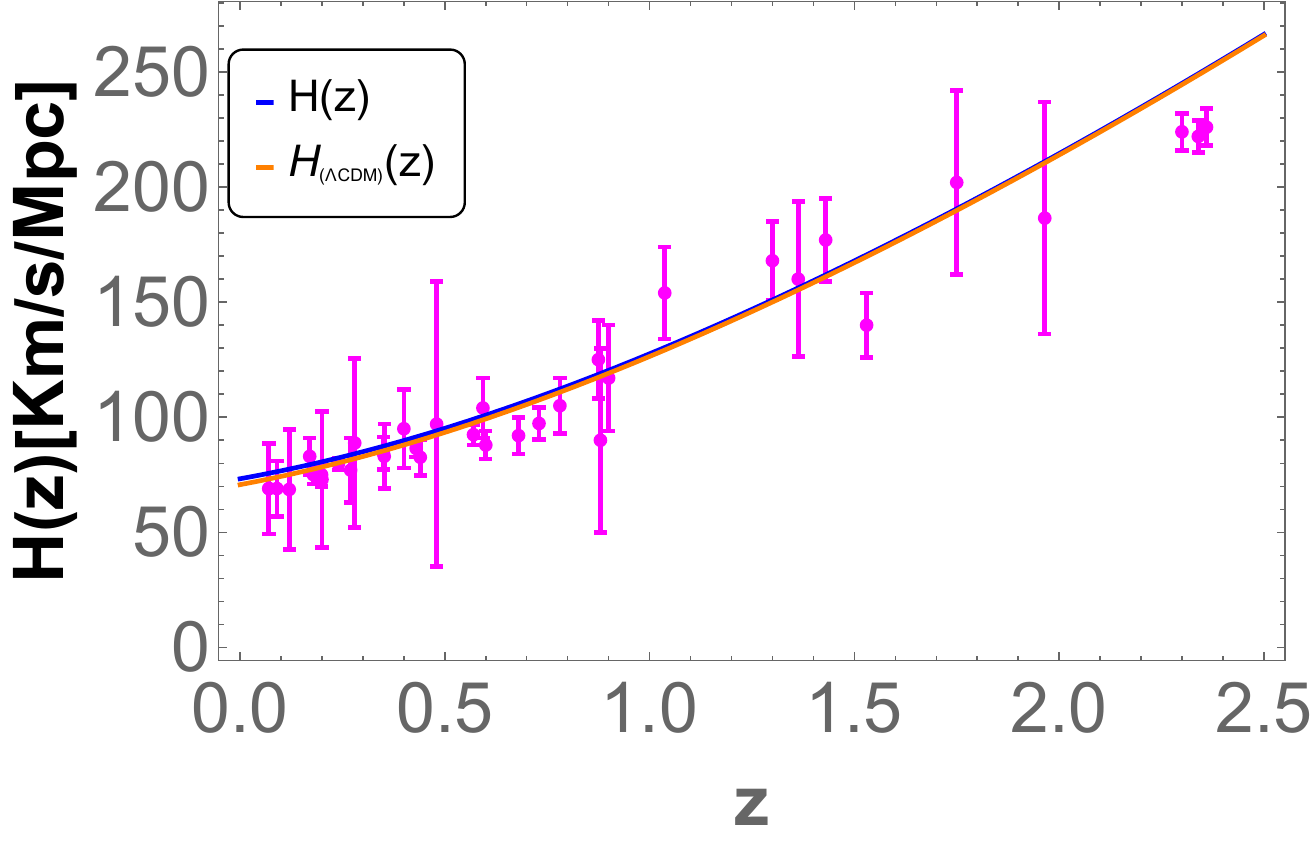}
\caption{Hubble parameter} \label{Hubblefig}
\end{subfigure}
\hspace{2.1cm}
\begin{subfigure}[h]{0.35\textwidth}
\centering
\includegraphics[width=78mm]{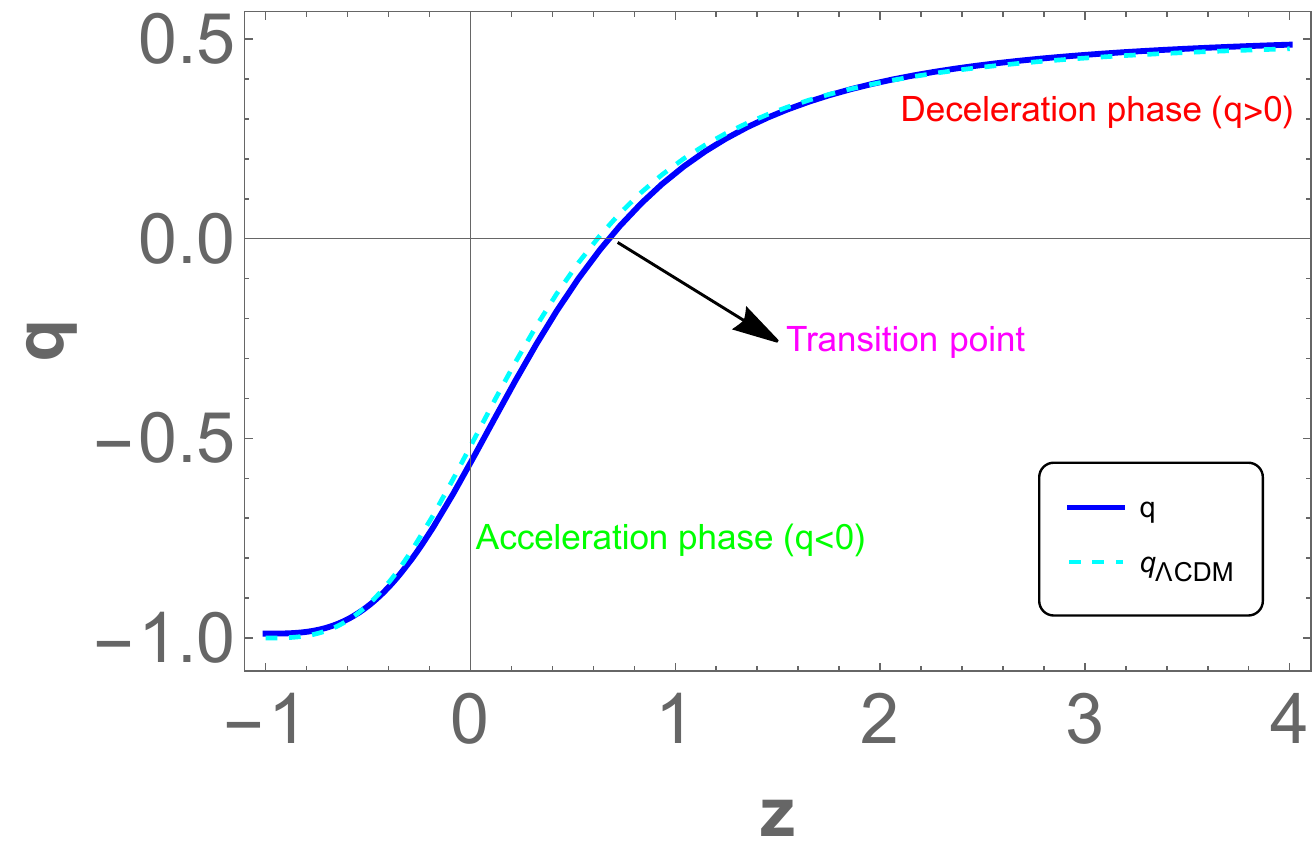}
\caption{Deceleration parameter}\label{decelerationparameterfig}
\end{subfigure}
\caption{The initial conditions are $x_C=10^{-4} ,\,y_C=10^{-6} ,\,u_C=10^{-8} ,\,\rho_C=0.883 ,\, n=2, \lambda=0.8, \sigma=12.2, \beta=14.05, Q=0.6$.}\label{h(z)q(z)m1}
\end{figure}

The numerical results were obtained using the ND-solve function in Mathematica. Our examination relied on the Hubble and Supernovae Ia (SNe Ia) observational data sets, which are described in detail in {\bf Appendix-\ref{appendix}}.
In Fig. \ref{Eosmfig}, a comparison of the equation of state (EoS) parameters for DE, total, and the EoS parameter for $\Lambda$CDM is presented. The graphs illustrate how $\omega_{tot}$, $\omega_{DE}$, and $\omega_{\Lambda CDM}$ tend towards $-1$ in the late stage of the Universe's evolution. The current value of $\omega_{DE}(z=0)\approx-1$,  $\omega_{tot}(z=0)\approx-0.68$ aligns with observations from the Planck Collaboration \cite{Aghanim:2018eyx}. Moreover, the energy density evolution for radiation, DE, and DM is depicted in Fig. \ref{densityparametersfig}. From these visuals, it is evident that radiation dominated phase was occured during the Universe's early phase. Furthermore, the graphs indicate that the DM-dominated phase is a relatively brief interval, culminating in the later appearance of the cosmological constant. The plot also shows that the contributions from DM ($\Omega_{m}\approx 0.3$) and DE ($\Omega_{DE}\approx 0.7$) which shows that these values are in agreement with the current cosmic observation studies. The era of matter-radiation equality occurs at around $z\approx 3387$, marked by a pointed arrow in Fig. \ref{densityparametersfig}. Fig. \ref{Hubblefig} presents the evolution of the Hubble rate alongside Hubble data points \cite{Moresco_2022_25}, with $H_{0} \approx 71$ Km/(Mpc sec) \cite{Aghanim:2018eyx}, indicating a close agreement with the standard $\Lambda$CDM model. The behavior of the deceleration parameter is analyzed in Figure \ref{decelerationparameterfig}, revealing transient behavior at $z\approx 0.66$ and showing consistency \cite{PhysRevD.90.044016a}. Currently, the deceleration parameter has a value of $q\approx -0.53$ \cite{PhysRevResearch.2.013028}. The development of the modulus function $\mu(z)$ is illustrated in Fig. \ref{luminositydistance}, demonstrating that the model's curve shows agreement with the $\Lambda$CDM model's modulus function $\mu_{\Lambda CDM}$, including data from 1048 Supernovae Ia (SNe Ia) refer {\bf Appendix-\ref{appendix}}.

\begin{figure}[H]
    \centering
\includegraphics[width=78mm]{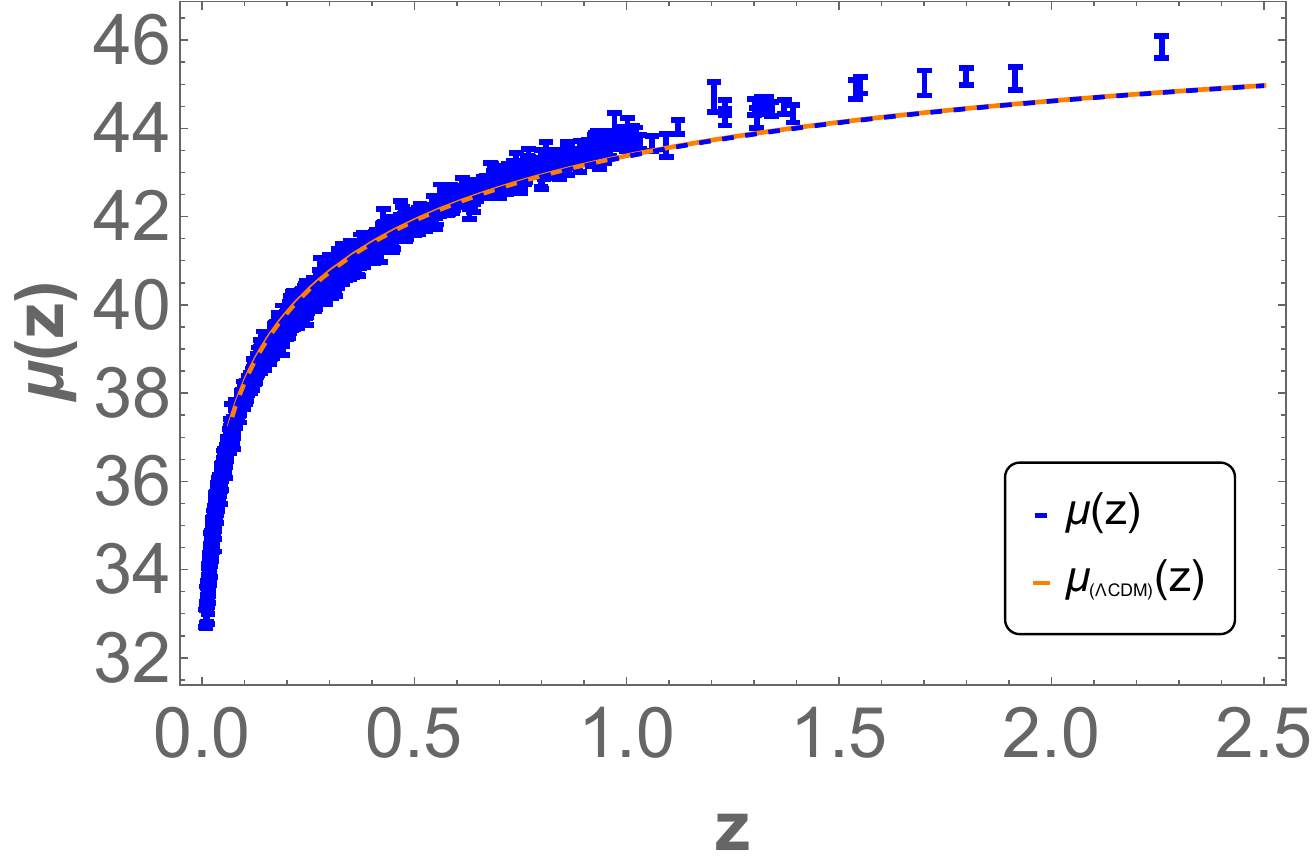}\caption{Plot for distance modulus function for $\mu(z)$ (observed) and $\mu_{\Lambda CDM}(z)$ (predicted). $x_C=10^{-4} ,\,y_C=10^{-6} ,\,u_C=10^{-8} ,\,\rho_C=0.883 ,\, n=2, \lambda=0.8, \sigma=12.2, \beta=14.05, Q=0.6$. } \label{mu(z)fig}
\end{figure}

\begin{figure}[H]
    \centering
\includegraphics[width=64mm]{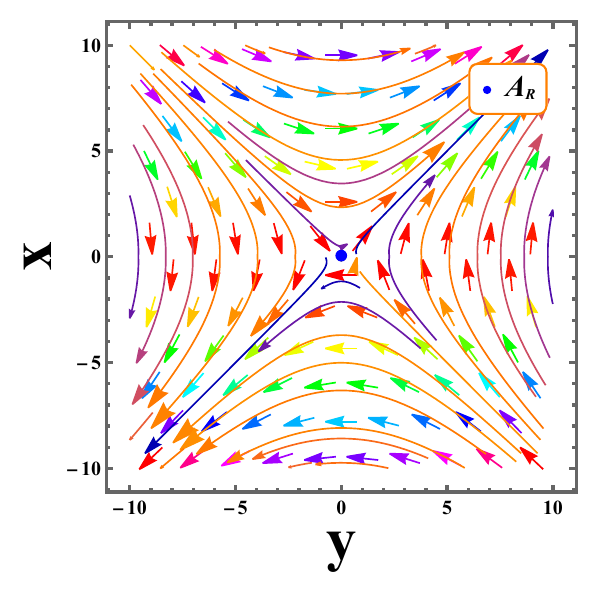}
\includegraphics[width=64mm]{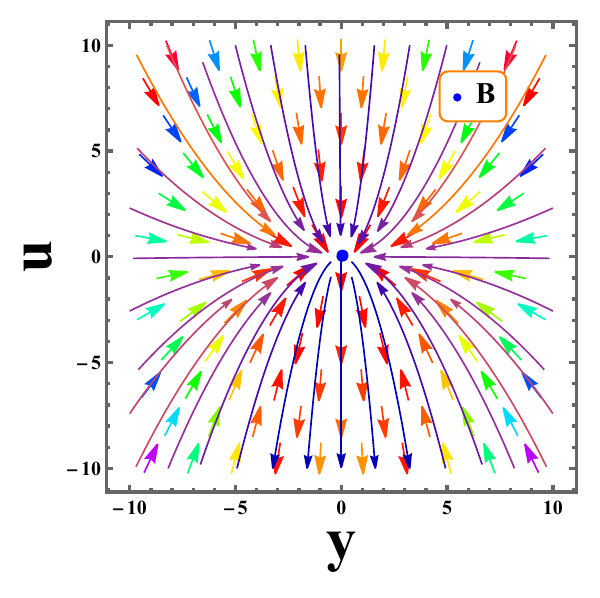}
\includegraphics[width=68mm]{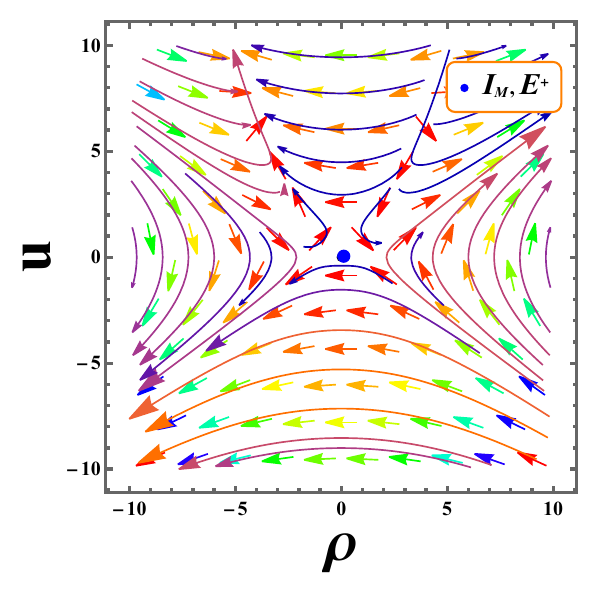}
\caption{2-D phase portrait for the critical point representing different epoch of Universe evolution.} \label{phaseportraitfig}
\end{figure}
To study and analyze the behaviour phase space trajectories at the critical points we obtain the 2-D phase space portrait diagram for the a critical points representing distinct epochs of the Universe evolution. The critical point $A_R$ representing a standard radiation-dominated epoch of the Universe evolution and phase space trajectories are showing saddle behaviour around this and same can be confirmed through the sign of the eigenvalues. The critical point $B$ representing de-Sitter solution is showing a attractor nature with all the phase space trajectories are showing attractor nature towards it. We have also plotted phase space plot for other two critical points which are representing matter dominated scaling solution $I_M$ and the stiff matter solution. These two critical points are showing saddle point behaviour, The phase space are plotted for the same values of the model parameters used in Fig. \ref{Eosdensitym1}.

\section{Summary and conclusion}\label{conclusion}

In this study, we have obtained a detailed analysis of an autonomous dynamical system with a general form of power law torsion coupling in the teleparallel scalar-tensor formalism. This study presents a extended version of the research conducted in the Sorkin-Schutz formalism \cite{Rodriguez_Benites_2025}. We succesfully formulate the autonomous dynamical system and studied the critical points representing different epochs of the Universe evolution. Also, we obtain detailed analysis of scaling solutions representing radiation $(C_{R}, J_{R})$, matter $(D_{M}, I_M)$ and the DE $(H)$. We analyse the stability of each of these critical point by obtaining the eigenvalues of each of these critical point. We constrain parameter $n$, at which the DE critical point $B$ is showing stable behaviour $n>1, \sigma>0$ is the suitable choice for parameter $n, \sigma$ where most of the above discussed critical points are showing stable behaviour.\\

To study the nature of phase space trajectories at critical points representing different epochs of the Universe evolution we plotted the phase plot in Fig. \ref{phaseportraitfig} for critical points $A_R, B, I_{M}, E^{+}$. The evolution plots presented in Fig. \ref{densityparametersfig}, describes Universe early to late time cosmic evolution phenomena. The plots are showing inagreement to describe the matter radiation equality at $z\approx 3387$ and at present the value of $\Omega_{m}\approx 0.3$ and $\Omega_{DE}\approx 0.7$. We have also plotted the behaviour of EoS parameters for $\omega_{de}, \omega_{tot}, \omega_{\Lambda}CDM$, which are inagreement with the $\Lambda$CDM model from early to the late time with $\omega_{DE}\approx -1$ \cite{Aghanim:2018eyx} at present time. The deceleration from early to late time of cosmic evolution show inagreement witht the $\Lambda$CDM model and can be visualised from Fig. \ref{decelerationparameterfig}. We have also plotted the error bar plots for the Hubble parameter using 31 data points from the Hubble data and the distance modulus function for our model, comparing with $\Lambda$CDM Ref Figs. \ref{Hubblefig}, \ref{mu(z)fig}. Both of these error bar plots are in agreement with the corresponding data sets with $H_{0}\approx 71$ Km/(Mpc sec) \cite{Aghanim:2018eyx}. This study can be extended further to analyse role of different parameters estimated here in the study of different cosmic phenomena like study of black holes, the neutron stars and the study of gravitational waves

\section{Appendix}\label{appendix}
\subsubsection*{Hubble data}
In this study, we consider 31 data points \cite{Moresco_2022_25} to illustrate the behavior of the Hubble rate for the above discussed scalar tensor gravity formalism. This model will then validated with the standard $\Lambda$CDM model. For which We should know that,

\begin{equation}\label{hubble_LCDM}
H_{\Lambda CDM}= H_{0}\sqrt{(1+z)^3 \Omega_{m}+(1+z)^4 \Omega_{r}+\Omega_{de}} \,,   
\end{equation}
\subsubsection*{\bf{Supernovae Ia}}
An additional element of our baseline dataset consists of the Pantheon collection of 1,048 SNIa distance measurements covering the redshift range of $0.01<z<2.3$ \cite{Scolnic_2018}. This dataset contains information from major observational projects, such as the Hubble Space Telescope (HST) survey, SDSS. By integrating information from various sources, the Pantheon dataset provides significant insights into the characteristics and behavior of Type Ia supernovae and their implications for the cosmos. Moreover, it illustrates the application of stellar luminosity for distance measurement in an expanding Universe, with the distance moduli function being a crucial aspect of this evaluation, expressed as follows:
\begin{equation}\label{panmoduli}
\mu(z_{i}, \Theta)=5 \log_{10}[D_{L}(z_i, \Theta)]+M    
\end{equation}
In this equation, $M$ represents the nuisance parameter, while $D_{L}$ signifies the luminosity distance. The luminosity distance can be computed using the following formula:
\begin{equation}\label{luminositydistance}
D(z_{i}, \Theta)=c (1+z_{i}) \int^{z_i}_{0} \frac{dz}{H(z, \Theta)} \,.   
\end{equation}
\bibliographystyle{utphys}
\bibliography{references}
\end{document}